# Prediction of Fishbone Linear Instability in Tokamaks with Machine Learning Methods

Z.Y. Liu [1], H.R. Qiu [2], G.Y. Fu [2*], Y. Xiao [2], Y.C. Chen [1], Z.J. Wang [1], Y.X. Wei [1†]

[1] Zhejiang Lab, Hangzhou, Zhejiang 311121, People's Republic of China

[2] Institute for Fusion Theory and Simulation and School of Physics, Zhejiang University, Hangzhou, Zhejiang 310027, People's Republic of China

* Email address for correspondence: gyfu@zju.edu.cn

† Email address for correspondence: yx_wei@zhejianglab.com

**Abstract**

A machine learning based surrogate model for fishbone linear instability in tokamaks is constructed. Hybrid simulations with the kinetic-magnetohydrodynamic (MHD) code M3D-K is used to generate the database of fishbone linear instability, through scanning the four key parameters which are thought to determine the fishbone physics. The four key parameters include (1) central total beta of both thermal plasma and fast ions, (2) the fast ion pressure fraction, (3) central value of safety factor $q$ and (4) the radius of $q = 1$ surface. Four machine learning methods including linear regression, support vector machines (SVM) with linear kernel, SVM with nonlinear kernel and multi-layer perceptron are used to predict the fishbone instability, growth rate and real frequency, mode structure respectively. Among the four methods, SVM with nonlinear kernel performs very well to predict the linear instability with accuracy $\approx 95\%$, growth rate and real frequency with $R^2 \approx 98\%$, mode structure with $R^2 \approx 98\%$.

## 1. Introduction

Energetic particles (EPs) can drive the MHD instabilities via wave-particle resonances in tokamaks, these instabilities evolve, then saturate and ultimately lead to EP transport, which is crucial to the performance of burning fusion plasmas [1]. The first observation of EP-driven mode is the fishbone instability in the Poloidal Divertor eXperiment (PDX) [2]. This experimental observation has drawn a lot of attention in the fusion community, because the fishbone instability can induce dramatic global EP transport [3]. The linear instability of fishbone was successfully explained by the resonant interaction between the $n = m = 1$ internal kink mode and the precession and/or transit frequencies of EPs [4-7], where $n/m$ represents the toroidal/poloidal number respectively. A number of self-consistent hybrid kinetic-MHD simulations were performed to study the linear and nonlinear physics of fishbone driven by neutral beam injection (NBI) [8-15], alpha particles [16], and energetic electrons [17]. Recently, the formations of internal transport barrier (ITB) accompanied with fishbone activity were studied in experiments [18-21] and simulations [22,23], regarding the shear flow generation through the nonlinear dynamics of fishbone [24,25].

Nowadays, machine learning methods show great potential to solve many scientific and engineering problems, with much better performance and efficiency comparing with traditional approaches. This trend is even more obvious as the computing power and capacity of modern computer clusters raise rapidly, especially for the Graphic Process Unit (GPU) development. Various machine learning algorithms have been applied in magnetic confinement fusion research, including the fast equilibrium solution [26-29], safety factor reconstruction [30,31], pedestal density prediction [32,33], plasma control [34-37] and disruption prediction [38-43]. It is also novel to use these new methods to identify and classify MHD instabilities and transport events in experiments [44-52], and to construct surrogate models for the first-principle simulations and transport calculations [53-58]. The machine learning methods can even solve the physical problems with the ability to merge the physical equations into the target loss functions, which are known as physics-informed neural networks (PINNs) [27, 59-63].

This work is aimed at constructing a machine learning based surrogate model for fishbone linear instability in tokamak. In order to generate the database of fishbone linear instability, we use the hybrid simulations with the kinetic-MHD code M3D-K, which has successfully simulated fishbone physics for different tokamaks [8,10,12-14,22]. There are many parameters that affect fishbone instability such as equilibrium configuration, $q$ profile, thermal pressure profile, and EP. Even though these one-dimensional profiles can be fitted as polynomial expressions, the dimension of parameter space is too large to build the database due to the time-consuming running of M3D-K code, for example, a linear case requires around 12 hours on average when running on clusters with 48 computing cores in parallel.

Due to the difficulty mentioned above, we restrict our problem to a much simpler one with several assumptions, (1) the circular cross section equilibrium of the HL-2A tokamak device size is adopted, (2) the thermal pressure and fast ion (here EP refers to fast ion induced by NBI) pressure profile are fixed with only variable central values, (3) the $q$ profile is monotonic and expressed as cubic polynomial with fixed boundary value, (4) we focus on the fishbone driven by trapped fast ions via toroidal procession frequency resonance. Within this reduced parameter space, four key parameters which are thought to determine the fishbone physics are selected and scanned to generate the database of fishbone linear instability. The four key parameters include (1) central total beta of both thermal plasma and fast ions, (2) the fast ion pressure fraction, (3) central value of $q$ and (4) the position of $q = 1$ surface. The resulting linear instability, growth rate and real frequency, mode structure as output, combined with the four parameters as featured input, can be easily collected to the standard data format for machine learning algorithms. Four machine learning methods including linear regression, SVM with linear kernel, SVM with nonlinear kernel and the simplest neural network multi-layer perceptron (MLP) are attempted, among them SVM with nonlinear kernel performs very well to predict the linear instability with accuracy $\approx 95\%$, growth rate and real frequency with $R^2 \approx 98\%$, mode structure with $R^2 \approx 98\%$. This is a great first step to verify and validate the efficiency of the surrogate model to predict fishbone linear instability, for the next step, the nonlinear evolution, saturation and fast ion transport can be explored, and the assumptions mentioned above can be dropped step by step via more advanced machine learning algorithms, eventually to construct a complete fishbone surrogate model independent of special devices.

The rest of paper are organized as follow. In section 2, a baseline case of fishbone linear instability on HL-2A tokamak configuration is displayed. On the basis of the baseline case, the four key parameters are scanned to generate the fishbone linear instability database which is described in section 3. In section 4, four machine learning methods including linear regression, SVM with linear kernel, SVM with nonlinear kernel and MLP are attempted to predict the linear instability in section 4.1, growth rate and real frequency in section 4.2, and mode structure in section 4.3 respectively. Finally, a summary is given in section 5.

## 2. Baseline case of fishbone on HL-2A tokamak

The kinetic-MHD code M3D-K is used to simulate fishbone linear instability on HL-2A tokamak driven by trapped fast ions via toroidal procession frequency resonance. M3D-K is a hybrid code in which the thermal plasma is treated by MHD model while the energetic particles are described by the drift-kinetic equation via the $\delta f$ particle-in-cell (PIC) method [8]. It has been extensively applied to study the interactions between MHD instabilities and EPs both in linear and nonlinear phase [10,12-14,22].

The basic equilibrium parameters are chosen based on HL-2A tokamak [64] with circular cross section equilibria, the major radius is $R_0 = 1.65$m, the minor radius is $a_0 = 0.4$m, magnetic field at magnetic axis is $B_0 = 1.3$T, central electron density is $n_{e0} = 2 \times 10^{19}/\text{m}^3$, the Alfvén speed is $v_A = 4.5 \times 10^6$m/s, the Alfvén frequency is $\omega_A = v_A/R_0 = 2.7 \times 10^6$/s, central total beta of both thermal plasma and fast ions is $\beta_{\text{total}} = 3\%$, the fast ion pressure fraction is $P_{\text{hot}}/P_{\text{total}} = 0.4$, the $q$ profile is $q = 0.8 + 3.2\psi^2$, the normalized thermal plasma pressure is $P_{\text{thermal}} = 1 - 2\psi^2 + \psi^4$. The injection energy of NBI is $E_0 = 45$keV, the fast ion pressure profile is $P_{\text{fast}} = P_{\text{hot}}\exp(-\langle\psi\rangle/\Delta\psi)$, here $\langle\psi\rangle$ is the normalized poloidal magnetic flux $\psi$ averaged over the particle orbit, $\Delta\psi = 0.2$, and use is made of a slowing down distribution as $f = [cH(v_0 - v)/(v^3 + v_c^3)]\exp[-(\Lambda - \Lambda_0)^2/\Delta\Lambda^2]$, where $c$ is a normalization factor, $H$ is the step function, $v_0 = \sqrt{2E_0/m_D}$ is the injection speed of NBI, $v_c = 0.79v_0$ is the critical velocity, $m_D$ is the mass of deuterium, $\Lambda \equiv \mu B_0/E$ is the pitch angle parameter, $\mu$ is the magnetic moment and $E$ is the energy of fast ions, and $\Lambda_0 = 1$, $\Delta\Lambda = 0.2$. The profiles of thermal plasma and fast ion pressure are shown in figure 1 with red lines, and $q$ profile is shown with blue line.

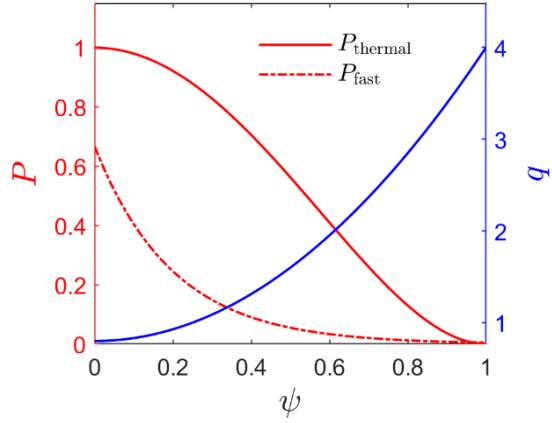

Figure 1. Thermal plasma and fast ion pressure profiles (red lines) and $q$ profile (blue line) used in the baseline case of fishbone instability.

The linear simulation results of the $n=1$ mode are displayed in figure 2, where the fast ion pressure fraction $P_{hot}/P_{total}$ ranges from 0 to 0.9 with an interval of 0.1, and the central total beta of both thermal plasma and fast ions is fixed at $\beta_{total}=3\%$. For zero fast ion pressure $P_{hot}/P_{total}=0$, the ideal $n=m=1$ internal kink mode is unstable with zero real frequency. For small values of $P_{hot}/P_{total}$, the effects of fast ions are stabilizing. When $P_{hot}/P_{total} \geq 0.2$, the fishbone instability driven by fast ions is excited [12], and its growth rate and finite real frequency both increase as $P_{hot}/P_{total}$ increases.

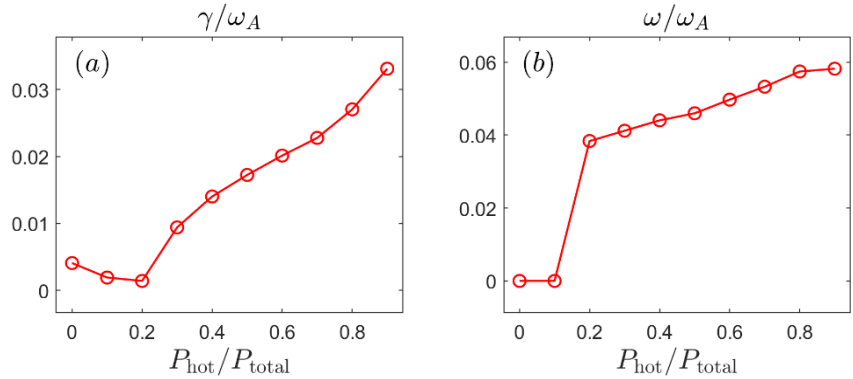

Figure 2. The (a) growth rate and (b) real frequency of the $n=1$ mode as a function of fast ion pressure fraction $P_{hot}/P_{total}$ with the central total beta fixed at $\beta_{total}=3\%$.

The mode structures for $P_{hot}/P_{total}=0$ and $P_{hot}/P_{total}=0.4$ are displayed in figure 3 respectively. Here we choose the output variable $U$, which is equivalent to the perturbed electric potential. With zero fast ion pressure, the mode is ideal MHD unstable with up-down symmetric structure in figure 3(a). While the fishbone mode structure shows a twisted feature in figure 3(b).

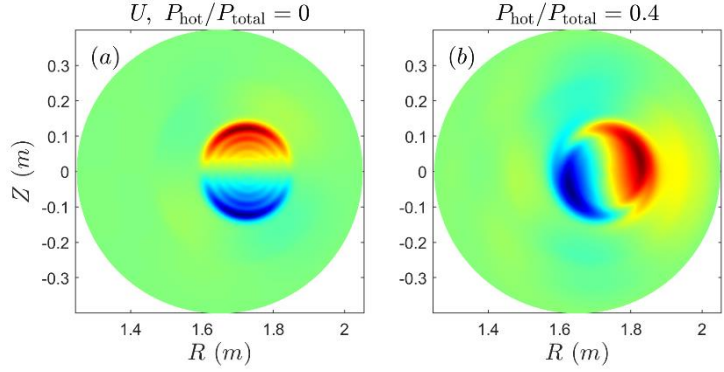

Figure 3. 2D mode structures for (a) $P_{\text{hot}}/P_{\text{total}} = 0$ and (b) $P_{\text{hot}}/P_{\text{total}} = 0.4$.

## 3. Fishbone linear instability database

As described in section 1, we generate the fishbone linear instability database through scanning the four key parameters including (1) central total beta of both thermal plasma and fast ions $\beta_{\text{total}}$, (2) the fast ion pressure fraction $P_{\text{hot}}/P_{\text{total}}$, (3) central value of $q$ profile denoted as $q_0$ and (4) the position of $q = 1$ surface $\psi|_{q=1}$.

The parameter range of $\beta_{\text{total}}$ ranges from 0.01 to 0.05 with an interval of 0.01, for $P_{\text{hot}}/P_{\text{total}}$, it ranges from 0 to 0.9 with an interval of 0.1. We adopt the cubic polynomial for $q$ as $q = q_0 + q_1\psi + q_2\psi^2 + q_3\psi^3$, where the coefficients satisfy $q_1 = 7.7 - 8q_0$, $q_0 + q_1 + q_2 + q_3 = 4$ such that the boundary $q$ is fixed at 4. As a result, the $q$ profile is only determined by $q_0$ and $\psi|_{q=1}$. The value of $q_0$ ranges from 0.7 to 0.95 with an interval of 0.05, and $\psi|_{q=1}$ ranges from 0.14 to 0.42 with an interval of 0.04. In order to consider some charge conditions with $q \geq 1$, we add three more $q_0$ values ranging from 1 to 1.1 with an interval of 0.05, and these $q$ profiles are simply profiles at $q_0 = 0.95$ adding a constant. Hence for $q_0 \geq 1$, $\psi|_{q=1}$ is just a parameter to determine the shape of profiles and do not represent the position of $q = 1$ surface. The $q$ profiles at $q_0 = 0.8$ but different $\psi|_{q=1}$ are illustrated in figure 4(a), and the $q$ profiles at $q_0 = 1.1$ are illustrated in figure 4(b) which are profiles at $q_0 = 0.95$ plus 0.15.

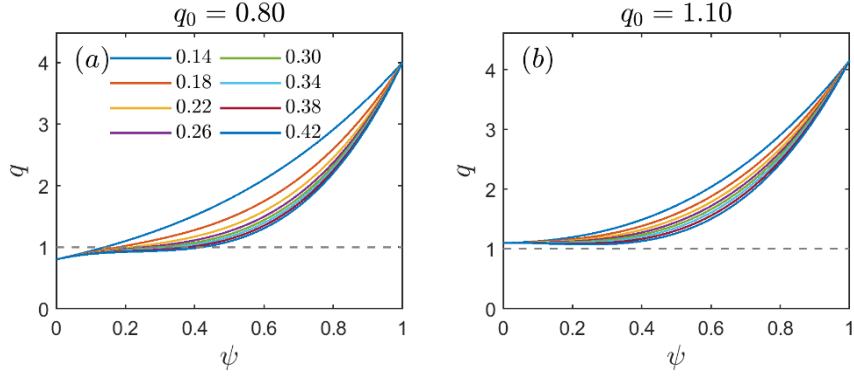

Figure 4. The $q$ profiles with (a) $q_0 = 0.8$ and different $\psi|_{q=1}$ ranging from 0.14 to 0.42, (b) $q_0 = 1.1$ and different values of $\psi|_{q=1}$.

To sum up, there are five $\beta_{\text{total}}$, ten $P_{\text{hot}}/P_{\text{total}}$, nine $q_0$ and eight $\psi|_{q=1}$ values to be scanned, the database contains totally 3600 parameter sets, which are used for M3D-K simulations in the Kirin No.2 cluster, the Tianhe No.3 supercomputer and Zhejiang Lab's clusters with 48 cores in parallel in each case. Each equilibrium is obtained self-consistently via VMEC code [65] by using the given $q$ profile and total pressure profile as input. In simulations, the time interval is set to be $0.01\tau_A$, where $\tau_A = 1/\omega_A$ is the Alfvén time, and the total simulation time is set to be $1000\tau_A$. However, we set a diagnosis module during each case run, to judge whether the growth rate and real frequency converge in time for every $50\tau_A$. The convergence criteria are that (1) the relative change of the mean growth rate and real frequency of the current time slice comparing to the prior time slice are both smaller than 1%, (2) the standard deviation of the growth rate and real frequency during the current $50\tau_A$ are both smaller than 1%, where the growth rate and real frequency are calculated by differentiating the amplitude and phase of the mode output from M3D-K code in adjacent time steps. If the two conditions are met simultaneously, the simulation will be terminated and move to the next case run automatically. Notice that the mode structure will be only output at the last time step of simulation to save the database storage space. The cases that do not converge are thought to be stable or marginal unstable modes, and they are not to be predicted in this work for two reasons, the first reason is that to obtain the correct growth rate and real frequency the simulation time should be set to be more than $1000\tau_A$ which needs much more computation resources, the second reason is that the stable or marginal unstable modes may not be observable in experiments, even though the marginal unstable modes can grow, the

equilibrium for such a long time $> 1000\tau_A$ may also be changed, which makes the simulation results meaningless.

Scanning growth rate and real frequency for different $\beta_{\text{total}}$ and $P_{\text{hot}}/P_{\text{total}}$ at $q_0 = 0.85$ and $\psi|_{q=1} = 0.26$ are displayed in figure 5, in which only the converged results are shown. One can see that as $P_{\text{hot}}/P_{\text{total}}$ exceeds a critical value and increases, the fishbones are destabilized and the growth rate and real frequency both rise, except for $\beta_{\text{total}} = 0.01$, the fishbone frequency decreases. As $\beta_{\text{total}}$ increases, the critical value of $P_{\text{hot}}/P_{\text{total}}$ for fishbone instability decreases. Results for other $q_0$ and $\psi|_{q=1}$ show qualitatively the same trend as in figure 5.

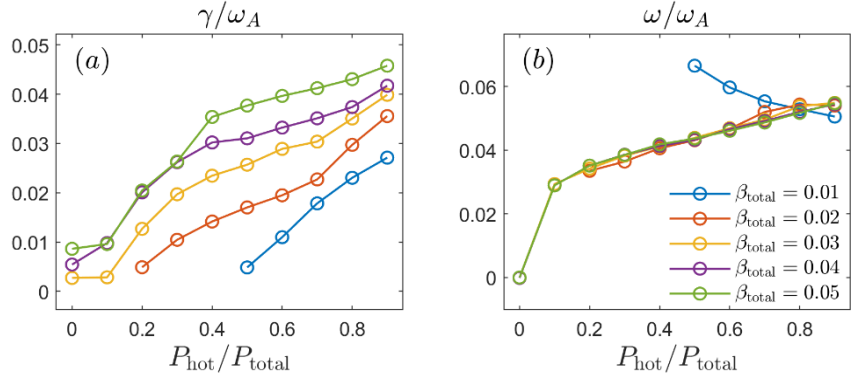

Figure 5. (a) Growth rate and (b) real frequency as a function of $P_{\text{hot}}/P_{\text{total}}$ with $q_0 = 0.85$, $\psi|_{q=1} = 0.26$ and different values of $\beta_{\text{total}}$.

Scanning growth rate and real frequency for different $q_0$ and $\psi|_{q=1}$ at $\beta_{\text{total}} = 0.03$ and $P_{\text{hot}}/P_{\text{total}} = 0.4$ are displayed in figure 6. There is an obvious discrepancy between the resonant fishbone at $q_0 < 1$ and the non-resonant fishbone at $q_0 \geq 1$. As $\psi|_{q=1}$ increases, the growth rate and real frequency of non-resonant fishbone increase and decrease respectively, while the growth rate of resonant fishbone first increases then decreases, and the fishbone frequency decreases continuously. The $q_0$ value also affects the results, for resonant fishbone, the larger $q_0$ results in larger growth rate. However, for non-resonant fishbone, the larger $q_0$ results in smaller growth rate and larger real frequency, and the effects of $q_0$ seem to be independent on $\psi|_{q=1}$, as shown by the three almost parallel lines in figure 6. Fishbones at other $\beta_{\text{total}}$ and $P_{\text{hot}}/P_{\text{total}}$ show qualitatively the same trend as in figure 6, but the internal kink modes behave differently and are not displayed here.

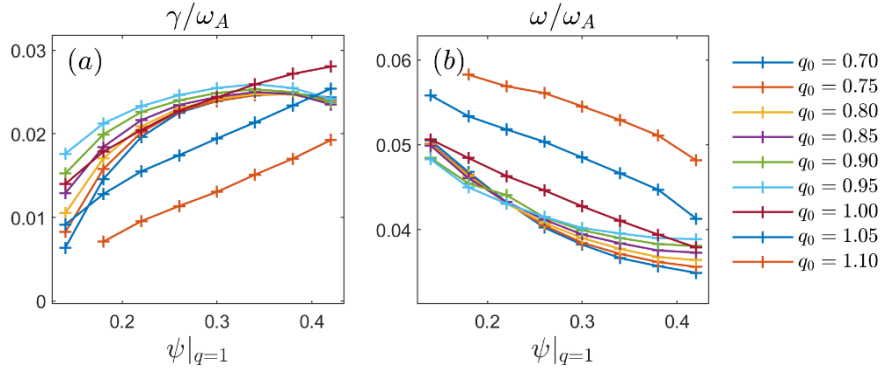

Figure 6. (a) Growth rate and (b) real frequency as a function of $\psi|_{q=1}$ with $\beta_{\text{total}} = 0.03$, $P_{\text{hot}}/P_{\text{total}} = 0.4$ and different values of $q_0$.

## 4. Performance of four machine learning methods

The resulting linear instability, growth rate and real frequency, mode structure as output, combined with the four parameters as featured input, can be easily collected to the standard data format for machine learning algorithms. The linear instability is featured as the number $c$, $c = 0$ for simulations that are not converged and $c = 1$ for converged according to the convergence criteria in section 3. The growth rate and real frequency are calculated as the mean value of the last $50\tau_A$ by differentiating the amplitude and phase of the mode. The mode structure predictions are more complicated. We first convert the two-dimensional (2D) mode structure into Fourier harmonics via fast Fourier transformation (FFT) in the poloidal direction, then only the two main components of $m = 1,2$ are considered. The last step is to convert the predicted $m = 1,2$ components into original cylindrical coordinate via inverse FFT.

Four machine learning methods including linear regression, SVM with linear kernel, SVM with nonlinear kernel and MLP are attempted to predict the linear instability in section 4.1, growth rate and real frequency in section 4.2, and mode structure in section 4.3 respectively. In this work, they are all carried out by using the scikit-learn library [66] with a personal computer with CPU of the 11th Gen Intel (R) Core (TM) i7-11700K @3.60GHz.

### 4.1. Linear instability

The linear instability prediction is a typically binary classification task. The four key parameters $\beta_{\text{total}}$, $P_{\text{hot}}/P_{\text{total}}$, $q_0$ and $\psi|_{q=1}$ of the 3600 samples are assigned to the feature matrix $X$, and the corresponding class labels of the convergence $c$ are assigned to the vector array $y$. Using the *train_test_split* function from scikit-learn's *model_selection* module, we randomly split the $X$ and $y$ arrays into 30% test and 70% train dataset, taking advantage of the built-in support for stratification via *stratify=y*.

Before predicting the linear instability, we assess feature importance of the four key parameters with an ensemble technique named random forest [67]. We train a forest of 500 trees on the resampled dataset and rank the four features by their respective importance measures via the *feature_importances_* attribute after fitting a *RandomForestClassifier* in scikit-learn library. Results are displayed in figure 7. One can see that the most important feature is $P_{\text{hot}}/P_{\text{total}}$, because there are critical values of $P_{\text{hot}}/P_{\text{total}}$ for the mode excitation. The second important feature is $\beta_{\text{total}}$, which determines the drive of the mode. The parameters $q_0$ and $\psi|_{q=1}$ are relatively less important for the mode instability.

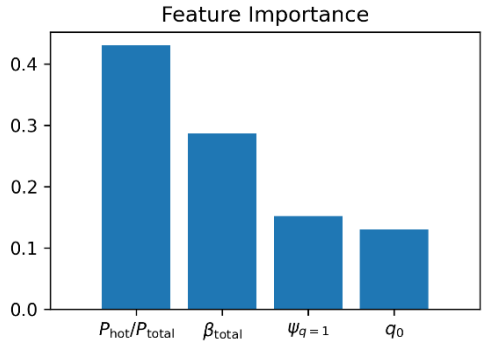

Figure 7. Feature importance of the four key parameters obtained by fitting a *RandomForestClassifier* containing 500 trees in scikit-learn library.

To predict the linear instability, we adopt four different machine learning methods which are listed in Table 1. These machine learning algorithms are easily implemented in the scikit-learn library via the streamlining workflows with pipelines [66]. Here, we combine the data-preprocessing transformer *StandardScaler* module to standardize the input features, and four estimators of *LogisticRegression*, *SVC* with kernal='linear', *SVC* with kernel='rbf' and *MLPClassifier* modules respectively into one pipeline to fit, predict and score in each method. The predicting accuracy of the four machine learning methods are listed in Table 1. The linear classification algorithms including Logistic regression and SVM with linear kernel can predict the linear instability with $\approx$ 90% accuracy both on train and test data set. However, the nonlinear algorithms including SVM with 'rbf' kernel and MLP perform well to predict the linear instability with $\approx$ 95% accuracy.

Table 1. The prediction accuracy by the four machine learning methods for train/test dataset.

| Train/Test | Logistic regression | Support vector machines (linear kernel) | Support vector machines (nonlinear kernel) | Multi-layer perceptron |
|---|---|---|---|---|
| Accuracy | 91.0/90.5% | 91.0/90.4% | 95.8/94.6% | 95.6/93.8% |

The hyper-parameters in these machine learning algorithms are tuning via *GridSearchCV* module defined in scikit-learn library to search for the optimal model [66]. Here, we carry out 10-fold cross-validation and compute the average accuracy across these 10-folds to assess the model performance. We set *n_jobs=-1* so that *GridSearchCV* can use all our 16 processing cores to speed up the grid search by fitting models to the different folds in parallel. The 10-fold cross-validations on the train dataset of the four machine learning methods result in accuracies of $91.1\% +/- 1.6\%$, $91.0\% +/- 1.4\%$, $95.2\% +/- 1.5\%$ and $95.2\% +/- 1.1\%$ respectively, where the number before/after $+/-$ represents mean value/standard deviation of the 10-folds. The small standard deviations indicate that these models are not under- or overfitted [67]. To further address those model issues, we use the learning curve to evaluate the model of SVM with 'rbf' kernel in figure 8 for illustration. Here, the blue and green line show the average accuracies from the returned cross-validated training and validation scores for the different sizes of the train dataset. Furthermore, we add the standard deviation of the accuracies to the plot as shown by the shaded area in the figure. As we can see in the learning curve, the SVM with 'rbf' kernel performs quite well on both the training and validation datasets if it has seen more than 1000 samples during training. We can also see that the training accuracy increases for train dataset with fewer than 250 samples, and the gap between validation and training accuracy widens, an indicator of an increasing degree of overfitting.

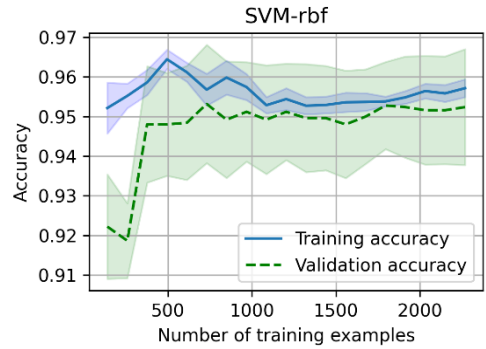

Figure 8. The learning curve of SVM with 'rbf' kernel.

The confusion matrix that lays out the performance of the model of SVM with 'rbf' kernel is displayed in figure 9. A confusion matrix is simply a square matrix that reports the counts of the true positive (TP), true negative (TN), false positive (FP), and false negative (FN) predictions of a classifier on the test dataset [67]. One can see in figure 9 that the model correctly classified 209 of the samples that belong to class $c = 0$ (TN) and 813 samples that belong to class $c = 1$ (TP) respectively. However, the model also incorrectly misclassified 25 samples from class $c = 1$ as class $c = 0$ (FN) and 33 samples from class $c = 0$ as class $c = 1$ (FP). These quantities can be used to calculate the true positive rate (TPR) and false positive rate (FPR), where $TPR = TP/(FN + TP)$ and $FPR = FP/(FP + TN)$.

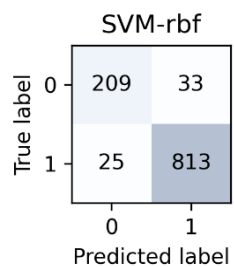

Figure 9. The confusion matrix of SVM with 'rbf' kernel.

The corresponding receiver operating characteristic (ROC) curves are displayed in figure 10. Based on the ROC curves, we can then compute the so-called ROC area under the curve (AUC) to characterize the performance of a classification model. A perfect classifier would fall into the top-left corner of the ROC curves with $TPR = 1$, $FPR = 0$, and $AUC = 1$. We can see in figure 10 that the nonlinear classifiers perform better than the linear classifiers.

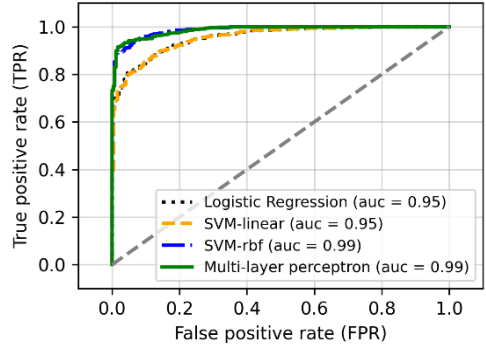

Figure 10. ROC curves of the four machine learning methods.

### 4.2. Growth rate and real frequency

Since our aim is to construct a surrogate model for fishbone, the first step is to separate the fishbone data from the database. Intuitively the most significant difference between the internal kink and fishbone is the real frequency, the former with zero frequency and the later with finite frequency as shown in figure 2. We give a scatterplot in the $\omega - \gamma$ space for all unstable modes ($c = 1$ class) in the database in figure 11, where $\gamma$ is the growth rate and $\omega$ is the real frequency. The corresponding histograms of $\gamma$ and $\omega$ are displayed in figure 11(b) and (c) respectively. The interesting results are that besides the $\omega = 0$ internal kink modes and finite frequency fishbones, there are several modes with moderate frequency $\omega \approx 0.01$. These modes are all obtained in the parameter region of $\beta_{\text{total}} = 0.04, 0.05$, $P_{\text{hot}}/P_{\text{total}} = 0.1$, $q_0 < 1$ and $\psi|_{q=1} = 0.42$, in which the internal kink and fishbone maybe coexist with comparable drive from the thermal plasma and fast ion pressure gradient respectively. In this work, we focus on the fishbone and simply separate the fishbone data with $\omega > 0.015$, i.e. points on the right of the red line in figure 11. The histogram of fishbone frequency is close to a gaussian distribution and its central value $\omega \approx 0.05$ as shown in figure 11(c), while the histogram of $\gamma$ is rather irregular in figure 11(b).

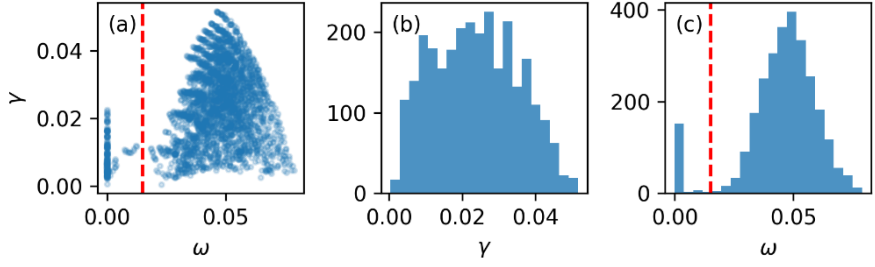

Figure 11. (a) Scatterplot in the $\omega - \gamma$ space, (b) histogram of $\gamma$ and (c) histogram of $\omega$ for all unstable modes ($c = 1$ class) in the database.

Let us look at relationships among the four key parameters and $\omega, \gamma$ using a correlation matrix in figure 12. Parameter $P_{\text{hot}}/P_{\text{total}}$ is the most important one to determine $\omega, \gamma$ with Pearson correlation coefficient $> 0.6$. However, the second important parameter is different, for $\gamma$ it is $\beta_{\text{total}}$, while for $\omega$ it is $\psi|_{q=1}$ with negative correlation. Notice that $q_0$ affects $\omega$ a lot, but affects little to $\gamma$.

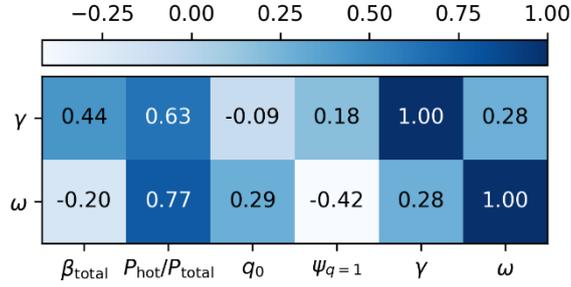

Figure 12. Correlation matrix among the four key parameters and $\omega, \gamma$.

The growth rate and real frequency prediction is a typically regression task. For this task, we randomly split the fishbone data into 30% test and 70% train dataset and adopt four different machine learning methods which are listed in Table 2. In contrast to the methods listed in Table 1, the linear method we used here is ridge regression, which is a L2-regularized linear regression where we simply add the squared sum of the weights to the mean squared error (MSE) loss function [67]. These machine learning algorithms are implemented via pipelines, in which we combine the data-preprocessing transformer *StandardScaler* module to standardize the input features and the four estimators of *Ridge*, *SVR* with kernal='linear', *SVR* with kernel='rbf' and

*MLPRegressor* respectively. The hyper-parameters in these machine learning algorithms have been tuned via *GridSearchCV* module. For regression tasks, the coefficient of determination $R^2$, which can be understood as a standardized version of the MSE, is used for better interpretability of the model's performance. If $R^2 = 1$, the model fits the data perfectly with a corresponding $MSE = 0$. Results in Table 2 show that the linear regression methods including Ridge regression and SVM with linear kernel can predict $\gamma$ with $R^2 \approx 85\%$ and $\omega$ with $R^2 \approx 80\%$. The MLP method performs not as good as in the classification task in section 4.1, but a little better than the linear methods. However, SVM with 'rbf' kernel shows a significantly stronger performance to predict $\gamma$ with $R^2 \approx 99\%$ and $\omega$ with $R^2 \approx 98\%$.

Table 2. The prediction $R^2$ of the growth rate and real frequency by the four machine learning methods for train/test dataset.

| Train/Test | Ridge regression | Support vector machines (linear kernel) | Support vector machines (nonlinear kernel) | Multi-layer perceptron |
|---|---|---|---|---|
| $\gamma$, $R^2$ | 85.6/84.4% | 85.5/84.2% | 99.6/99.3% | 86.9/86.2% |
| $\omega$, $R^2$ | 80.6/80.7% | 79.9/79.7% | 98.9/98.1% | 84.4/85.3% |

In figure 13, learning curve to evaluate the model of SVM with 'rbf' kernel is displayed. Here, we carry out 10-fold cross-validation and compute the average $R^2$ across these 10-folds to assess the model performance. As we can see in the figure, the SVM with 'rbf' kernel performs quite well on both the training and validation datasets if it has seen more than 1000 samples during training. There is no obvious under- or overfitting problem.

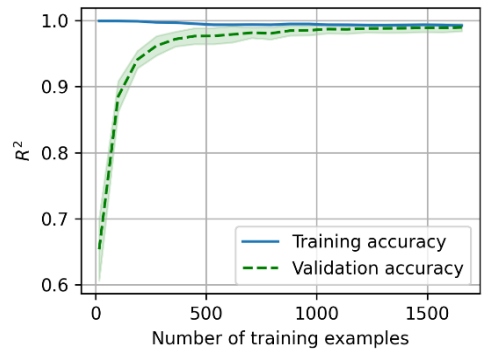

Figure 13. The learning curve of SVM with 'rbf' kernel.

The regression results can be further evaluated by the predict plot to evaluate the model of SVM with 'rbf' kernel in figure 14. In the case of a perfect prediction, these points would be exactly in the black line. In this practical application, they are randomly scattered around the black line, which indicates that the regression by the model of SVM with 'rbf' kernel is pretty good.

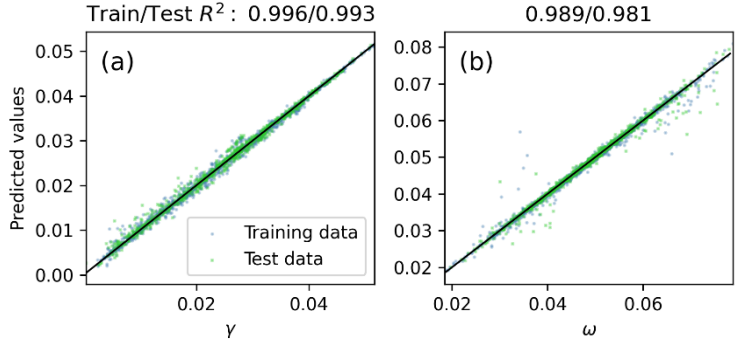

Figure 14. Prediction plot for (a) growth rate and (b) real frequency for SVM with 'rbf' kernel.

## 4.3 Mode structure

The mode structure predictions are rather challenged, because the data size is pretty large for one case, for example, there are 20201 points of the output variable $U$ in the R-Z plane in figure 3. Meanwhile, predicting mode structures in the cylindrical coordinates directly leads to poor performance. For these reasons, we convert the mode structures into Fourier harmonics via FFT in the poloidal direction. The equilibrium used in M3D-K code is generated by VMEC code and recorded as poloidal magnetic flux $\psi(R,Z)$. We first normalized $\psi$ by $\psi \to (\psi - \psi_{\min})/(\psi_{\max} - \psi_{\min})$, where $\psi_{\min}$ and $\psi_{\max}$ are the minimum and maximum of $\psi$. As a result, $\psi = 0$ on the magnetic axis and $\psi = 1$ on the boundary. The poloidal angles are defined as $\theta \equiv \text{atan}((Z - Z_{\text{axis}})/(R - R_{\text{axis}}))$, where $(R_{\text{axis}}, Z_{\text{axis}})$ is the position of the magnetic axis. A structured grid of $\psi_i = 0:0.001:1$, $\theta_j = 0:2\pi/200:2\pi$ is defined to calculate the coordinate $(R_{i,j}, Z_{i,j})$ on $(\psi_i, \theta_j)$, and then $U_{i,j}$ by using the cubic interpolation technique to the output variables $\psi(R,Z)$ and $U(R,Z)$ from M3D-K code. Once the discrete points of $U(\psi, \theta)$ are known, we apply FFT in $\theta$ direction to get various harmonics $U_m$, $m = 1,2 ...$, and their amplitudes and phases are denoted as $|U_m|$ and $\arg(U_m)$ respectively. In figure 15, we display the transform process of one case with the parameters of $\beta_{total} = 0.03$, $P_{hot}/P_{total} = 0.4$, $q_0 = 0.85$ and $\psi|_{q=1} = 0.26$. The 2D mode structure of this case is shown in figure 15(a), corresponding amplitudes and phases of $m = 1,2$ are shown in figure 15(c) and (d). The mode structure in figure 15(b) is constructed by inverse FFT, but only the two main components of $m = 1,2$ are retained. One can see that for fishbone, the $m = 1$ component is most important and much larger than the $m = 2$ component in amplitudes.

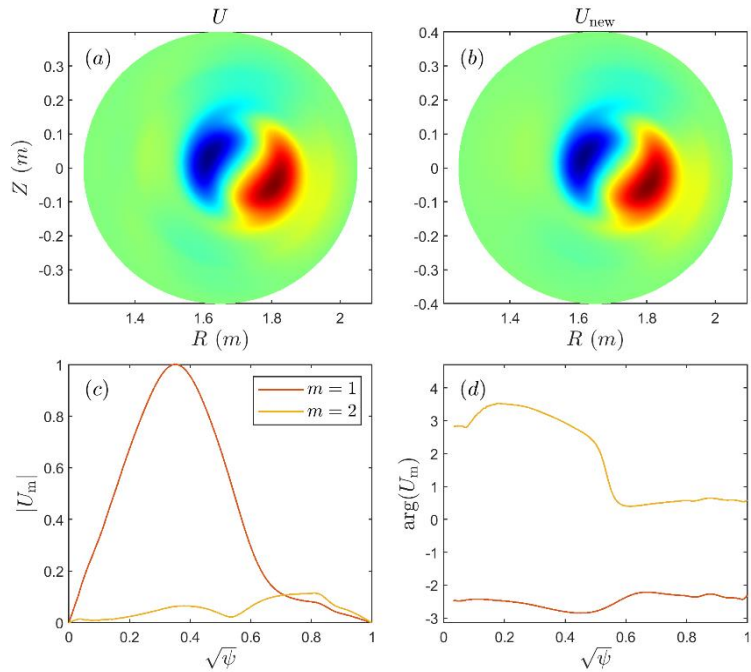

Figure 15. The 2D mode structure of (a) one case with the parameters of $\beta_{total} = 0.03$, $P_{hot}/P_{total} = 0.4$, $q_0 = 0.85$ and $\psi|_{q=1} = 0.26$, (b) the reconstruction by inverse FFT using the (c) amplitudes and (d) phases of the two main components $m = 1,2$.

After implementing the transformation mentioned above, the 2D mode structures are converted into four 1D ones for each case. Notice that there is a random phase for the mode in linear stage, we resize the phases by subtracting a constant from them and making $\arg(U_m)$ of $m = 1$ vanish as $\psi \to 0$. We also take advantage of the criteria $\omega > 0.015$ used in section 4.2 to separate the fishbone data from the database and randomly split the fishbone data into 30% test and 70% train dataset. This is still a huge regression task with totally 4004 targets for 1001 points in $\psi$ coordinate. However, it is convenient to adopt the *MultiOutputRegressor* module in scikit-learn library, and *n_jobs=-1* is set to speed up fitting by using all our 16 processing cores [66]. Applying the four machine learning methods listed in Table 3, which are the same as in section 4.2, we predict the amplitudes and phases of the two main components $m = 1,2$, and then convert them back into R-Z plane to calculate the predicting $R^2$. Results are listed in Table 3, one can see that the ridge regression and MLP can predict the 2D mode structure with $R^2 \approx 85\%$ and $R^2 \approx 91\%$, SVM with linear kernel gains a better performance with $R^2 \approx 95\%$. Again, the model of SVM with 'rbf' kernel performs even better with $R^2 \approx 98\%$.

Table 3. The prediction $R^2$ of the 2D mode structure by the four machine learning methods for train/test dataset.

| Train/Test | Ridge regression | Support vector machines (linear kernel) | Support vector machines (nonlinear kernel) | Multi-layer perceptron |
|---|---|---|---|---|
| $U$, $R^2$ | 85.7/85.6% | 95.4/95.3% | 98.1/98.1% | 91.1/90.9% |

The prediction plot to evaluate the model of SVM with 'rbf' kernel is displayed in figure 16, where the subplots show the histogram of $\Delta U = U_{\text{predicted}} - U$. We can see that most points are aligned along the black line.

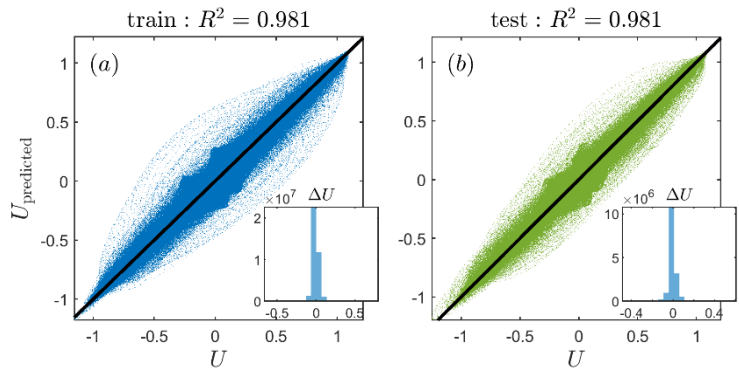

Figure 16. Prediction plot for the 2D mode structures of (a) train dataset and (b) test dataset to evaluate the model of SVM with 'rbf' kernel, where the subplots show the histogram of $\Delta U = U_{\text{predicted}} - U$.

When the fishbones are located very near the magnetic axis, which indeed happens for small $\psi|_{q=1}$, the mode structure $U(R,Z)$ approaches zero at most positions in the R-Z plane. In these cases, $R^2$ may not be a good score to evaluate the model's performance. Here, we use two image-relevant figures of merit, the peak signal-to-noise ratio PSNR [68] and mean structural similarity SSIM [69]. The PSNR is originally developed to estimate the degree of artifacts due to image compression compared to an original image, and the SSIM is used to estimate perceptual similarity (or perceived differences) between the true and reproduced images based on the inter-dependence of adjacent spatial pixels in the images [26]. The histograms of PSNR and SSIM for the 2D mode structures of the train and test dataset are displayed in figure 17. One can see that most PSNR values locate around 35dB, and most SSIM values are close to unity, indicating that the predicated 2D mode structures preserve the original quality of the simulated ones with a reasonable degree.

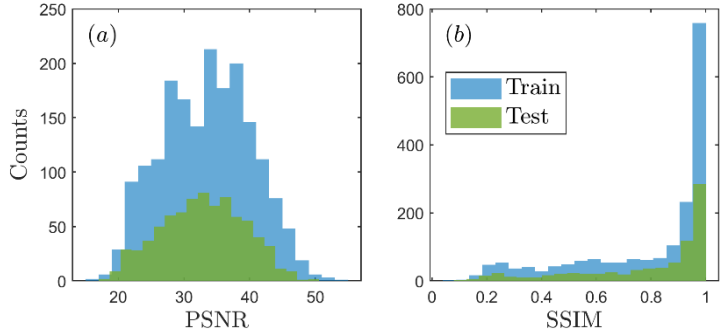

Figure 17. The histograms of (a) PSNR and (b) SSIM for the 2D mode structures of the train and test dataset to evaluate the model of SVM with 'rbf' kernel.

In figure 18, one case is selected from the test dataset randomly, and the simulated and predicted (by SVM with 'rbf' kernel) results of the amplitudes and phases of the two main components $m = 1,2$ are displayed by the solid and dashed lines respectively, the parameters for this case are $\beta_{\text{total}} = 0.02$, $P_{\text{hot}}/P_{\text{total}} = 0.4$, $q_0 = 0.8$ and $\psi|_{q=1} = 0.34$. We can see that the amplitudes and $m = 1$ phases are predicted quite well, while the $m = 2$ phases are predicted roughly. In fact, for most cases in the test dataset, the $m = 1$ amplitudes and phases can be predicted with high accuracy, predictions of the $m = 2$ components are slightly worse, probably due to their small amplitudes.

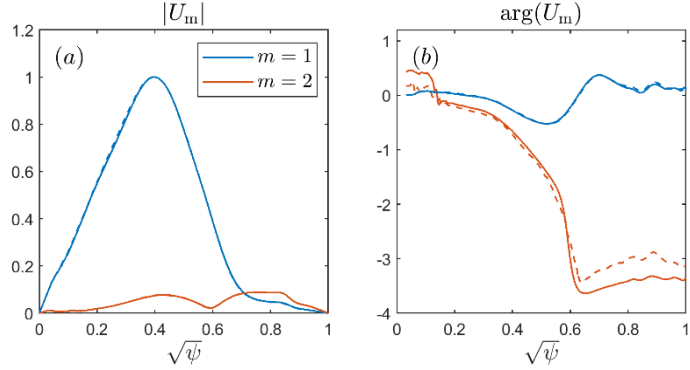

Figure 18. The simulated (solid lines) and predicted (dashed lines, by SVM with 'rbf' kernel) results of the (a) amplitudes and (b) phases of the two main components $m = 1,2$, the parameters for this case are $\beta_{\text{total}} = 0.02$, $P_{\text{hot}}/P_{\text{total}} = 0.4$, $q_0 = 0.8$ and $\psi|_{q=1} = 0.34$.

Nevertheless, the 2D mode structure can be captured mostly as shown in figure 19, because the main component of fishbone is $m = 1$.

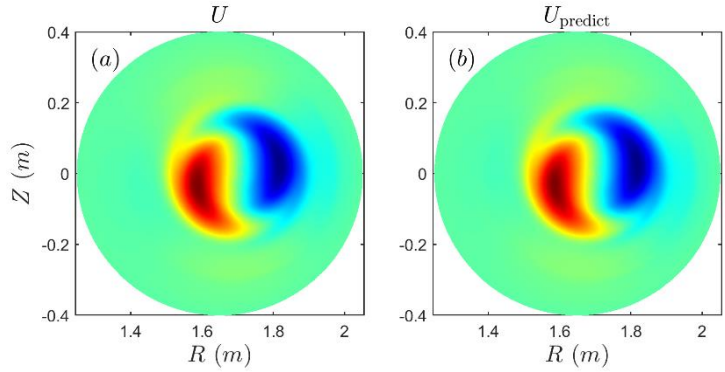

Figure 19. (a) The simulated and (b) the predicted (by SVM with 'rbf' kernel) 2D mode structure corresponding to the case in figure 18.

The average predicting time for one case by SVM with 'rbf' kernel consumed in the personal computer is $7.7 \times 10^{-6}$s for the linear instability prediction, $4.0 \times 10^{-5}$s for the growth rate and real frequency prediction and $1.8 \times 10^{-2}$s for the mode structure prediction. The first two predictions meet the requirement of real-time control system of tokamaks with $\approx 1\text{ms}$ time intervals. The mode structure prediction is a little more time-consuming, however, it may be reduced by applying the algorithm in GPU devices. These results have shown the strong potential of machine learning methods to surrogate the first-principle simulations with a much faster speed.

## 5. Summary

In summary, we have constructed a machine learning based surrogate model for fishbone linear instability in tokamaks. A relatively simple scenario is considered with circular cross section equilibria of HL-2A tokamak device size, fixed thermal pressure and fast ion pressure profile with variable central values, monotonic $q$ profile with fixed boundary value and assumption that fishbones are driven by trapped fast ions via toroidal procession frequency resonance. Under a typical set of parameters, the baseline case is simulated via M3D-K code in section 2, where the internal kink modes and fishbones are identified depending on the fast ion pressure fraction $P_\text{hot}/P_\text{total}$, with central total beta of both thermal plasma and fast ions $\beta_\text{total}$ fixed. The four key parameters including $\beta_\text{total}$, $P_\text{hot}/P_\text{total}$, $q_0$ and $\psi|_{q=1}$ are scanned to set up the database of the fishbone linear instability containing totally 3600 parameter sets in section 3. Taking advantage of the powerful scikit-learn library, four machine learning methods including linear regression, SVM with linear kernel, SVM with nonlinear kernel and MLP are attempted to predict the linear instability, growth rate and real frequency, and mode structure. For the linear instability prediction, the predicting accuracy for train and test dataset are almost equal and there is no obvious under- or overfitting. Among the four machine learning methods, SVM with nonlinear kernel performs best with accuracy $\approx 95\%$. For the growth rate and real frequency prediction, we separate the fishbone data with $\omega > 0.015$ from the database. In this case, SVM with nonlinear kernel shows strongest performance with $R^2 \approx 98\%$, much better than the other three methods. For the mode structure prediction, we first convert the 2D mode structures into Fourier harmonics via FFT in the poloidal direction, and then only retain $m = 1,2$ components in amplitudes and phases. The predicted $m = 1,2$ components are then converted back into original cylindrical coordinate via inverse FFT to calculate the predicting $R^2$. Again, the model of SVM with nonlinear kernel performs best with $R^2 \approx 98\%$. The two image-relevant figures of merit, PSNR and SSIM are also evaluated, and the predicated 2D mode structures reproduce the simulated ones with an acceptable quality.

The machine learning methods used in this work are simple, but predict the linear instability, the growth rate and real frequency, mode structure of fishbone with good performance, especially with the model of SVM with nonlinear kernel. The reason is that we take advantage of the four key parameters as featured input, as a result, they can be easily collected to the standard data format for traditional machine learning algorithms such as Logistic regression, SVM and MLP. The advantage of this scheme is that it is easier to generate the database, and the four key parameters can be quickly accessed in experiments and inputted into our models. If the 1D profiles of $q$, the thermal plasma and fast ion pressure reconstructed from experimental data are taken as featured input, then some advanced machine learning algorithms, for example, the deep learning with convolutional neural network [55], should be used. For the next step, the nonlinear evolution, saturation and fast ion transport induced by fishbones can be explored, more complex and realistic equilibrium configurations and 1D profiles can be taken into account as featured input, eventually to construct a complete fishbone surrogate model independent of special devices, and meet the requirement of the real-time control system of tokamaks.

## Acknowledgment

M3D-K runs in this work are carried out on the Kirin No.2 (http://ifts.zju.edu.cn/about/?155.html?lg=en) at Zhejiang University, the Tianhe No.3 (https://www.nscc-tj.cn/cjjs_zy_th3) and Zhejiang Lab's (http://alkaidos.org.cn) high-performance clusters.

This work is supported by the National MCF Energy R&D Program of China (Grant No. 2019YFE03050001) and Zhejiang Lab under the project of Digital twin system for intelligent simulation and control of nuclear fusion (124000-AC2304).

## References

[1] Chen L. and Zonca F. 2016 Physics of Alfvén waves and energetic particles in burning plasmas, *Rev. Mod. Phys.* **88** 015008

[2] McGuire K. et al 1983 Study of high-beta magnetohydrodynamic modes and fast-ion losses in PDX, *Phys. Rev. Lett.* **51** 1925–1925

[3] White R. B., Goldston R. J., McGuire K., Boozer A. H., Monticello D. A., and Park W. 1983 Theory of mode-induced beam particle loss in tokamaks, *Phys. Fluids* **26**, 2958

[4] Chen L., White R.B. and Rosenbluth M.N. 1984 Excitation of internal kink modes by trapped energetic beam ions, *Phys. Rev. Lett.* **52** 1122–5

[5] Coppi B. and Porcelli F. 1986 Theoretical model of fishbone oscillations in magnetically confined plasmas, *Phys. Rev. Lett.* **57** 2272–5

[6] Betti R. and Freidberg J.P. 1993 Destabilization of the internal kink by energetic-circulating ions, *Phys. Rev. Lett.* **70** 3428–30

[7] Porcelli F., Stankiewicz R., Kerner W. and Berk H.L. 1994 Solution of the drift-kinetic equation for global plasma modes and finite particle orbit widths, *Phys. Plasmas* **1** 470–80

[8] Fu G.Y., Park W., Strauss H.R., Breslau J., Chen J., Jardin S. and Sugiyama L.E. 2006 Global hybrid simulations of energetic particle effects on the n=1 mode in tokamaks: Internal kink and fishbone instability, *Phys. Plasmas* **13** 052517

[9] Kim, C.C. 2008 Impact of velocity space distribution on hybrid kinetic-magnetohydrodynamic simulation of the (1,1) mode, *Phys. Plasmas* **15** 7

[10] Wang F., Fu G.Y. and Shen W. 2016 Nonlinear fishbone dynamics in spherical tokamaks, *Nucl. Fusion* **57** 016034

[11] Pei Y., Xiang N., Hu Y., Todo Y., Li G., Shen W. and Xu. L. 2017 Kinetic-MHD hybrid simulation of fishbone modes excited by fast ions on the experimental advanced superconducting tokamak (EAST), *Phys. Plasmas* **24** 032507

[12] Shen W., Wang F., Fu G.Y., Xu L., Li G. and Liu C. 2017 Hybrid simulation of fishbone instabilities in the EAST tokamak, *Nucl. Fusion* **57** 116035

[13] Ren, Z.Z., *et. al.* 2018 Hybrid simulations of fishbone instabilities and Alfven eigenmodes in DIII-D tokamak, *Phys. Plasmas* **25** 12

[14] Shen W., *et. al.*, 2020 Hybrid simulation of fishbone instabilities with reversed safety factor profile, *Nucl. Fusion* **60** 10

[15] Liu C., *et. al.*, 2022 Thermal ion kinetic effects and Landau damping in fishbone modes, *J. Plasma Phys.* **88** 6

[16] Brochard G., Dumont R., Lütjens H., Garbet X., Nicolas T. and Maget P. 2020 Nonlinear dynamics of the fishbone-induced alpha transport on ITER, *Nucl. Fusion* **60** 126019

[17] Vlad G., Briguglio S., Fogaccia G., Zonca F., Fusco V. and Wang X. 2013 Electron fishbone simulations in tokamak equilibria using XHMGC, *Nucl. Fusion* **53** 083008

[18] Gunter S., Gude A., Hobirk J., Maraschek M., Saarelma S., Schade S., Wolf R.C., ASDEX Upgrade Team 2001 MHD phenomena in advanced scenarios on ASDEX Upgrade and the influence of localized electron heating and current drive, *Nucl. Fusion* **41** 1283

[19] Liu, Z.X. *et al* 2020 Experimental observation and simulation analysis of the relationship between the fishbone and ITB formation on EAST tokamak, *Nucl. Fusion* **60** 122001

[20] Zhang B. *et al* 2022 Progress on physics understanding of improved confinement with fishbone instability at low q95<3.5 operation regime in EAST, *Nucl. Fusion* **62** 126064

[21] He X. X. *et al* 2022 The ITB dynamics controlled by internal kink modes on HL-2A tokamak, *Plasma Phys. Control. Fusion* **64** 015007

[22] Ge W., Wang Z. X., Wang F., Liu Z., and Xu L. 2022 Multiple interactions between fishbone instabilities and internal transport barriers in EAST plasmas, *Nucl. Fusion* **63** 016007

[23] Brochard G. *et al* 2024 Saturation of fishbone instability by self-generated zonal flows in tokamak plasmas, *Phys. Rev. Lett.* **132** 075101

[24] Pinches S.D. *et al* 2021 Fishbone Generation of Sheared Flows and the Creation of Transport Barriers, ECA Vol. 25A, 57–60

[25] Liu Z.Y. and Fu G.Y. 2023 A Simple Model for Internal Transport Barrier Induced by Fishbone in Tokamak Plasmas, *J. Plasma Phys.* **89** 905890612

[26] Joung S., Kim J., Kwak S., Bak J.G., Lee S.G., Han H.S., Kim H.S., Lee G., Kwon D. and Ghim Y.-C. 2020 Deep neural network Grad–Shafranov solver constrained with measured magnetic signals, *Nucl. Fusion* **60** 016034

[27] Kaltsas D. A., Throumoulopoulos G. N. 2022 Neural network tokamak equilibria with incompressible flows, *Phys. Plasmas* **29** 022506

[28] Lao L. L. *et al* 2022 Application of machine learning and artificial intelligence to extend EFIT equilibrium reconstruction, *Plasma Phys. Control. Fusion* **64** 074001

[29] Lu J., Hu Y., Xiang N., and Sun Y. 2023 Fast equilibrium reconstruction by deep learning on EAST tokamak, *AIP Advances* **13** 075007

[30] Wakatsuki T., Suzuki T., Hayashi N., Oyama N. and Ide S. 2019 Safety factor profile control with reduced central solenoid flux consumption during plasma current ramp-up phase using a reinforcement learning technique, *Nucl. Fusion* **59** 066022

[31] Wei X., Sun S., Tang W., Lin Z., Du H. and Dong G. 2023 Reconstruction of tokamak plasma safety factor profile using deep learning, *Nucl. Fusion* **63** 086020

[32] Kit A., Järvinen A. E., Frassinetti L., Wiesen S. and JET Contributors 2023 Supervised learning approaches to modeling pedestal density, *Plasma Phys. Control. Fusion* **65** 045003

[33] Gillgren A., Fransson E., Yadykin D., Frassinetti L., Strand P. and JET Contributors 2022 Enabling adaptive pedestals in predictive transport simulations using neural networks, *Nucl. Fusion* **62** 096006

[34] Degrave J. *et al* 2022 Magnetic control of tokamak plasmas through deep reinforcement learning, *Nature* **602** 414–419

[35] Wan C., Yu Z., Wang F., Liu X. and Li J. 2021 Experiment data-driven modeling of tokamak discharge in EAST, *Nucl. Fusion* **61** 066015

[36] Wan C., Yu Z., Pau A., Sauter O., Liu X., Yuan Q. and Li J. 2023 A machine-learning-based tool for last closed-flux surface reconstruction on tokamaks, *Nucl. Fusion* **63** 056019

[37] Dubbioso S., De Tommasi G., Mele A., Tartaglione G., Ariola M., Pironti A. 2023 A Deep Reinforcement Learning approach for Vertical Stabilization of tokamak plasmas, *Fusion Engineering and Design* **194** 113725

[38] Kates-Harbeck J., Svyatkovskiy A. and Tang W. 2019 Predicting disruptive instabilities in controlled fusion plasmas through deep learning, *Nature* **568** 526–531

[39] Fu Y., Eldon D., Erickson K., Kleijwegt K., Lupin-Jimenez L., Boyer M. D., Eidietis N., Barbour N., Izacard O. and Kolemen E. 2020 Machine learning control for disruption and tearing mode avoidance, *Phys. Plasmas* **27** 022501

[40] Zheng W. *et al* 2022 Overview of machine learning applications in fusion plasma experiments on J-TEXT tokamak, *Plasma Sci. Technol.* **24** 124003

[41] Zhu J.X., Rea C., Granetz R.S., Marmar E.S., Sweeney R., Montes K. and Tinguely R.A. 2023 Integrated deep learning framework for unstable event identification and disruption prediction of tokamak plasmas, *Nucl. Fusion* **63** 046009

[42] Shen C., Zheng W., Ding Y., Ai X., Xue F., Zhong Y., Wang N., Gao L., Chen Z., Yang Z., Chen Z., Pan Y. and J-TEXT team 2023 IDP-PGFE: an interpretable disruption predictor based on physics-guided feature extraction, *Nucl. Fusion* **63** 046024

[43] Zheng W. *et al* 2023 Disruption prediction for future tokamaks using parameter-based transfer learning, *Communications Physics* **6** 181

[44] Smith D. R., Fonck R. J., McKee G. R., Diallo A., Kaye S. M., LeBlanc B. P., Sabbagh S. A. 2016 Evolution patterns and parameter regimes in edge localized modes on the National Spherical Torus Experiment, *Plasma Phys. Control. Fusion* **58** 045003

[45] Bustos A., Ascasíbar E., Cappa A. and Mayo-García R. 2021 Automatic identification of MHD modes in magnetic fluctuation spectrograms using deep learning techniques, *Plasma Phys. Control. Fusion* **63** 095001

[46] Kaptanoglu A. A. *et al* 2022 Exploring data-driven models for spatiotemporally local classification of Alfven eigenmodes, *Nucl. Fusion* **62** 106014

[47] Wei Y., Levesque J. P., Hansen C., Mauel M. E. and Navratil G. A. 2023 MHD mode tracking using high-speed cameras and deep learning, *Plasma Phys. Control. Fusion* **65** 074002

[48] Lee J. E., Seo P. H., Bak J. G. and Yun G. S. 2021 A machine learning approach to identify the universality of solitary perturbations accompanying boundary bursts in magnetized toroidal plasmas, *Scientific Reports* **11** 3662

[49] Han W., Pietersen R. A., Villamor-Lora R., Beveridge M., Offeddu N., Golfinopoulos T., Theiler C., Terry J. L., Marmar E. S., Drori I. 2022 Tracking blobs in the turbulent edge plasma of a tokamak fusion device, *Scientific Reports* **12** 18142

[50] Han W., Offeddu N., Golfinopoulos T., Theiler C., Terry J.L., Wüthrich C., Galassi D., Colandrea C., Marmar E.S. and the TCV Team 2023 Estimating cross-field particle transport at the outer midplane of TCV by tracking filaments with machine learning, *Nucl. Fusion* **63** 076025

[51] Salazar L., Heuraux S., Sabot R., Krämer-Flecken A., Tore Supra Team and TEXTOR Team 2022 Extraction of quasi-coherent modes based on reflectometry data, *Plasma Phys. Control. Fusion* **64** 104007

[52] Yang K.N. et al 2024 Neural network identification of the weakly coherent mode in I-mode discharge on EAST, *Nucl. Fusion* **64** 016035

[53] Piccione A., Berkery J.W., Sabbagh S.A. and Andreopoulos Y. 2020 Physics-guided machine learning approaches to predict the ideal stability properties of fusion plasmas, *Nucl. Fusion* **60** 046033

[54] Liu Y., Lao L., Li L. and Turnbull A. D. 2020 Neural network based prediction of no-wall βN limits due to ideal external kink instabilities, *Plasma Phys. Control. Fusion* **62** 045001

[55] Dong G., Wei X., Bao J., Brochard G., Lin Z. and Tang W. 2021 Deep learning based surrogate models for first-principles global simulations of fusion plasmas, *Nucl. Fusion* **61** 126061

[56] Li H., Fu Y., Li J. and Wang Z. 2021 Machine learning of turbulent transport in fusion plasmas with neural network, *Plasma Sci. Technol.* **23** 115102

[57] Van Mulders S., Felici F., Sauter O., Citrin J., Ho A., Marin M. and van de Plassche K.L. 2021 Rapid optimization of stationary tokamak plasmas in RAPTOR: demonstration for the ITER hybrid scenario with neural network surrogate transport model QLKNN, *Nucl. Fusion* **61** 086019

[58] Clement M.D., Logan N.C. and Boyer M.D. 2022 Neoclassical toroidal viscosity torque prediction via deep learning, *Nucl. Fusion* **62** 026022

[59] Heinonen R. A. and Diamond P. H. 2020 Turbulence model reduction by deep learning, *Phys. Rev. E* **101** 061201

[60] Mathews A., Francisquez M., W. Hughes J., Hatch D. R., Zhu B., and Rogers B. N. 2021 Uncovering turbulent plasma dynamics via deep learning from partial observations, *Phys. Rev. E* **104** 025205

[61] Mathews A., Hughes J. W., Terry J. L., and Baek S. G. 2022 Deep Electric Field Predictions by Drift-Reduced Braginskii Theory with Plasma-Neutral Interactions Based on Experimental Images of Boundary Turbulence, *Phys. Rev. Lett.* **129** 235002

[62] Poels Y., Derks G., Westerhof E., Minartz K., Wiesen S. and Menkovski V. 2023 Fast dynamic 1D simulation of divertor plasmas with neural PDE surrogates, *Nucl. Fusion* **63** 126012

[63] Rossi R., Gelfusa M., and Murari A. on behalf of JET contributors 2023 On the potential of physics-informed neural networks to solve inverse problems in tokamaks, *Nucl. Fusion* **63** 126059

[64] Duan X.R. et al 2009 Overview of experimental results on HL-2A, *Nucl. Fusion* **49** 104012

[65] Hirschman S. P. and Whitson J. C. 1983 Steepest-descent moment method for three-dimensional magnetohydrodynamic equilibria, *Phys. Fluids* **26** 3553–3568

[66] scikit-learn developers 2023 https://scikit-learn.org/stable/index.html

[67] Raschka S., Liu Y. and Mirjalili V. 2022 Machine Learning with PyTorch and Scikit-Learn, Packt Publishing

[68] Huynh-Thu Q. and Ghanbari M. 2008 Scope of validity of PSNR in image/video quality assessment, *Electr. Lett.* **44** 800

[69] Wang Z., Bovik A.C., Sheikh H.R. and Simoncelli E.P. 2004 Image quality assessment: from error visibility to structural similarity, *IEEE Trans. Image Process.* **13** 600